\documentclass[sigconf]{acmart}
\usepackage{listings}
\usepackage{xcolor}

\colorlet{punct}{red!60!black}
\definecolor{background}{HTML}{EEEEEE}
\definecolor{delim}{RGB}{20,105,176}
\colorlet{numb}{magenta!60!black}

\definecolor{lightgray}{rgb}{.9,.9,.9}
\definecolor{darkgray}{rgb}{.4,.4,.4}
\definecolor{purple}{rgb}{0.65, 0.12, 0.82}

\lstdefinelanguage{json}{
    basicstyle=\normalfont\ttfamily,
    numbers=left,
    numberstyle=\scriptsize,
    stepnumber=1,
    numbersep=8pt,
    showstringspaces=false,
    breaklines=true,
    frame=lines,
    backgroundcolor=\color{background},
    literate=
     *{0}{{{\color{numb}0}}}{1}
      {1}{{{\color{numb}1}}}{1}
      {2}{{{\color{numb}2}}}{1}
      {3}{{{\color{numb}3}}}{1}
      {4}{{{\color{numb}4}}}{1}
      {5}{{{\color{numb}5}}}{1}
      {6}{{{\color{numb}6}}}{1}
      {7}{{{\color{numb}7}}}{1}
      {8}{{{\color{numb}8}}}{1}
      {9}{{{\color{numb}9}}}{1}
      {:}{{{\color{punct}{:}}}}{1}
      {,}{{{\color{punct}{,}}}}{1}
      {\{}{{{\color{delim}{\{}}}}{1}
      {\}}{{{\color{delim}{\}}}}}{1}
      {[}{{{\color{delim}{[}}}}{1}
      {]}{{{\color{delim}{]}}}}{1},
      morestring=[b]',
}

\lstdefinelanguage{js}{
    basicstyle=\normalfont\ttfamily,
    numbers=left,
    numberstyle=\scriptsize,
    stepnumber=1,
    numbersep=8pt,
    showstringspaces=false,
    breaklines=true,
    frame=lines,
    backgroundcolor=\color{background},
    stringstyle=\color{red}\ttfamily,
    morestring=[b]',
    morestring=[b]",
     keywords={typeof, new, true, false, catch, function, return, null, catch, switch, var, let, if, in, while, do, else, case, break},
  keywordstyle=\color{blue}\bfseries,
  ndkeywords={class, export, boolean, throw, implements, import, this},
  ndkeywordstyle=\color{darkgray}\bfseries,
  identifierstyle=\color{black}
}

\lstdefinelanguage{JavaScript}{
  keywords={typeof, new, true, false, catch, function, return, null, catch, switch, var, if, in, while, do, else, case, break},
  keywordstyle=\color{blue}\bfseries,
  ndkeywords={class, export, boolean, throw, implements, import, this},
  ndkeywordstyle=\color{darkgray}\bfseries,
  identifierstyle=\color{black},
  sensitive=false,
  comment=[l]{//},
  morecomment=[s]{/*}{*/},
  commentstyle=\color{purple}\ttfamily,
  stringstyle=\color{red}\ttfamily,
  morestring=[b]',
  morestring=[b]"
}

\makeatletter
\renewcommand\@formatdoi[1]{\ignorespaces}
\makeatother

\AtBeginDocument{%
  \providecommand\BibTeX{{%
    \normalfont B\kern-0.5em{\scshape i\kern-0.25em b}\kern-0.8em\TeX}}}

\setcopyright{none}
\copyrightyear{Accepted at Connected World \& Semantic Interoperability Workshop 2022}
\acmYear{}
\acmConference[CWSI '22]{Connected World \& Semantic Interoperability Workshop 2022}{November 07, 2022}{Delft, Netherlands}
\acmISBN{}

\begin{document}

\title{Applying the Web of Things Abstraction to Bluetooth Low Energy Communication}

\author{Michael Freund}
\orcid{1234-5678-9012}
\affiliation{%
  \institution{Fraunhofer Institute for Integraded Circuits IIS}
  \city{Nürnberg}
  \country{Germany}
}
\email{michael.freund@iis.fraunhofer.de}

\author{Rene Dorsch}
\orcid{1234-5678-9012}
\affiliation{%
  \institution{Fraunhofer Institute for Integraded Circuits IIS}
  \city{Nürnberg}
  \country{Germany}
}
\email{rene.dorsch@iis.fraunhofer.de}

%\author{Thomas Wehr}
%\affiliation{%
%  \institution{Fraunhofer Institute for Integraded Circuits IIS}
%  \city{Nürnberg}
%  \country{Germany}
%}

\author{Andreas Harth}
\affiliation{%
  \institution{Fraunhofer Institute for Integraded Circuits IIS}
  \institution{Friedrich-Alexander-Universität Erlangen Nürnberg}
  \city{Nürnberg}
  \country{Germany}
}

\renewcommand{\shortauthors}{M. Freund, et al.}

\begin{abstract}
We apply the Web of Things (WoT) communication pattern, i.e., the semantic description of metadata and interaction affordances, to Internet of Things (IoT) devices that rely on non-IP-based protocols, using Bluetooth Low Energy (LE) as an example. The reference implementation of the WoT Scripting API, node-wot, currently supports only IP-based application layer protocols such as HTTP and MQTT. However, a significant number of IoT devices do not communicate over IP, but via other network layer protocols, e.g. L2CAP used by Bluetooth LE. To leverage the WoT abstraction in Bluetooth Low Energy communication, we specified two ontologies to describe the capabilities of Bluetooth LE devices and transmitted binary data, considered the different interaction possibilities with the Linux Bluetooth stack BlueZ, and due to better documentation, used the D-Bus API to implement Bluetooth LE bindings in JavaScript. Finally, we evaluated the latencies of the bindings in comparison to the BlueZ tool \texttt{bluetoothctl}, showing that the Bluetooth LE bindings are on average about 16 percent slower than the comparison program during connection establishment and about 6 percent slower when disconnecting, but have almost the same performance during reading (about 3 percent slower).
\end{abstract}




\maketitle

\section{Introduction}
% Was bietet IoT?
%The interconnection of network-enabled devices forms the foundation of the Internet of Things (IoT). The large-scale deployment of IoT promises unprecedented increases in productivity, higher product quality as well as efficiency \cite{I4Benefits} in industrial environments, and primarily increases to the quality of life in consumer environments \cite{8058258}, especially in the smart home sector.

% Was ist das Problem mit IoT? -> Interoperabilität -> Warum ist schwer zu erreichen?
%In the current IoT environment, there are many different proposed standards from multiple organizations that are often incompatible among each other \cite{syed2021iot}.  A missing reference standard leads to a lack of interoperability across the entire industry, which prevents users from leveraging the full potential the IoT offers and also creates major technical and business challenges \cite{aloi2017enabling}. The interoperability challenges in the IoT include, for instance, the integration of new devices into existing ecosystems or the development of applications that interact with a wide range of IoT devices and platforms provided by different manufacturers.
%In general, a distinction can be made between several types of interoperability. For example, if technical interoperability is achieved, devices can exchange bytes with each other. If syntactic interoperability is reached, there is a uniform structure and syntax of the transmitted data, and if the meaning of the data is also described, semantic interoperability is established.

The Internet of Things (IoT) is a heterogeneous domain, which makes it difficult to establish interoperability between all connected devices. One of the challenges is that IoT devices communicate via a variety of different connectivity protocols and transmit data in different formats \cite{rahman2020comprehensive}. 
%In some cases, existing standards are also misused, which further complicates the achievement of interoperability.

% Bestehender Ansatz -> WoT -> beschreibt mit TD
With the Web of Things (WoT) architecture \cite{wotArchitektur}, the World Wide Web Consortium (W3C) has presented a potential strategy for achieving semantic interoperability in the IoT. The distinctive aspect of the WoT architecture is that no new standard is created that IoT devices have to implement, but rather transmitted data and interaction affordances of already existing devices are described semantically in a uniform way that can be read by humans and machines. The WoT ecosystem also specifies a software stack, the so-called WoT Scripting API \cite{ScriptingAPI}, which makes it simpler for programmers to create applications involving various IoT devices. The reference open source implementation of the WoT Scripting API is the JavaScript library node-wot\footnote{\url{https://github.com/eclipse/thingweb.node-wot}}.

%There are several approaches to accomplish semantic interoperability between IoT devices, like the Asset Administration Shell (AAS) \cite{AAS}, which was developed alongside the Industry 4.0 concept, the Digital Twin Definition Language (DTDL) \cite{dtdl}, which is part of Microsoft's product portfolio, and the Web of Things (WoT) architecture \cite{wotArchitektur} developed by the World Wide Web Consortium (W3C). Compared to the other approaches mentioned, the WoT architecture has the advantage that it does not define a new API for accessing devices, but instead describes devices and their interaction affordances semantically \cite{app10186519} in a uniform way that is readable by both humans and machines with the help of a so-called Thing Description (TD) \cite{TD}.

% Was ist WoT und worauf ist es fokusiert?
The WoT focuses especially on the protocols located in the application layer of the internet protocol suite\footnote{\url{https://datatracker.ietf.org/doc/html/rfc1122}} such as HTTP, MQTT and Websockets. The high-level communication standards of the application layer use transport layer protocols like the connection-oriented TCP or the connectionless UDP to transfer data between a destination and a source. TCP and UDP require the internet layer protocol, or network layer protocol in the OSI model, IP to transfer packages to other networks. The IP protocol interacts via the link layer with the MAC-based connectivity protocols WiFi and Ethernet.

% Problem bei WoT
However, the IoT sector does not only consist of devices that implement the internet protocol suite and rely on WiFi or Ethernet connectivity in order to connect directly to the web but also of a large number of devices based on other connectivity standards, such as Bluetooth Low Energy (LE), ZigBee or Z-Wave \cite{7460395}. Bluetooth LE, for example, is according to the 2020 Eclipse Developer Survey\footnote{\url{https://iot.eclipse.org/community/resources/iot-surveys/}} the third most relevant connectivity protocol in the IoT area only surpassed by WiFi and Ethernet.

Since node-wot is focused on protocols that are based on IP, it is currently only possible to use the WoT abstraction and the associated advantages in a subset of all available IoT devices.

We aim to show that the communication patterns presented by the WoT architecture, i.e., the semantic description of devices and the communication model based on interaction affordances, are also suitable for non-IP-based connectivity protocols, which we will demonstrate using Bluetooth LE and the GATT protocol as an example.  

The contributions presented in this paper are
\begin{itemize}
 \item the design of ontologies to describe Bluetooth Low Energy devices and transmitted binary data,
 \item the application of the WoT patterns to Bluetooth Low Energy and GATT,
 \item the implementation and evaluation of Bluetooth Low Energy bindings and Thing Descriptions.
\end{itemize}

We will first give an example of the advantages that the application of WoT patterns on Bluetooth LE provides for programmers (section \ref{section:Example}). After that we will discuss related work (section \ref{section:RelatedWork}) and the background of Bluetooth, with focus on Bluetooth Low Energy (section \ref{section:BackgroundOfBLE}). We define two vocabularies, one describing Bluetooth LE devices and the other describing the transmitted binary data, and map the abstract WoT operations to the concrete GATT methods (section \ref{section:BLEforWoT}). We then explore the different ways of interacting with the Bluetooth protocol stack in Linux operating systems (section \ref{section:BLEonLinux}) and implement and evaluate protocol bindings with three example Thing Descriptions for node-wot (section \ref{section:Implementation} and \ref{section:Evaluation}). Finally, we conclude our work and give an outlook on the future (section \ref{section:conclusion}).

\section{Bluetooth LE using WoT}
\label{section:Example}
We want to demonstrate the advantages of the WoT abstraction with an example. We use a Bluetooth LE controlled lamp, that can be switched on or off and the current power status can be read. In the following, we consider the steps a programmer has to perform to read the power status, once by direct interaction with the lamp and once by using the WoT abstraction. 

The first approach, without WoT abstraction, is shown in listing \ref{ohne}. To read the power status, a programmer must first start the scan process and search for the IoT device using the MAC address (line 3). Once the device is found, the programmer can connect, access the GATT server, and select the appropriate service and characteristic (lines 4 to 8). The programmer is at this step able to read the bytes of the power characteristic (line 9). But the read bytes need to be decoded to get the actual value (line 10). Without the WoT abstraction, a programmer must know the MAC address, the IDs of the service and the characteristic, as well as the format of the transmitted binary data to access the power status.

\begin{lstlisting}[float, language=js, firstnumber=1, caption=Example pseudocode program reading the power status of the Bluetooth LE lamp without WoT abstraction., captionpos=b, label=ohne]
adapter = defaultAdapter();
adapter.startDiscovery();
device = adapter.getDevice('BE:58:30:00:CC:11');
device.connect();
adapter.stopDiscovery();
gattSrv = device.gatt();
service = gattSrv.getService('0000fff0-0000-1000-8000-00805f9b34fb');
char = service.getCharacteristic('0000fff3-0000-1000-8000-00805f9b34fb');
buffer = char.readValue();
status = buffer.readIntLE(offset=0, bytelength=1);
device.disconnect();
\end{lstlisting}

The WoT abstraction, on the other hand, uses a so-called Thing Description (TD) \cite{TD}. The TD is a JSON-LD document that semantically describes the metadata and affordances of a device (a WoT Thing or Thing). The affordances of a Thing are formally grouped into three categories. Writeable and readable attributes are assigned to \textit{properties}, executable and long-running processes to \textit{actions}, and asynchronous notifications to \textit{events}. Each affordance has a corresponding \textit{forms} field in which protocol-specific information such as permitted operations, method names, and a Uniform Resource Identifier\footnote{\url{https://www.rfc-editor.org/rfc/rfc3986.html}} (URI) are defined.

To read the power status of the Bluetooth LE lamp using the WoT abstraction (listing \ref{lst:mit}), a programmer simply needs to parse the TD (line 1), connect to the device and execute the abstract WoT method \texttt{readproperty} with parameter 'power' (lines 2 and 3).
The interaction information, like MAC address, service ID, characteristic ID, and metadata of transmitted bytes are described in the TD. The programmer only needs to know the name of the property, which is in this case 'power'. The mapping of the abstract WoT operation (e.g. \texttt{readproperty}) to corresponding protocol methods (e.g. \texttt{read}) is performed by so-called WoT protocol bindings \cite{BINDINGTemplate}.

The interaction with the BLE lamp using the WoT abstraction is much easier to realize for a programmer since a large part of the required information is already described in the TD.

\begin{lstlisting}[float, language=js, firstnumber=1, caption=Example pseudocode program reading the power status of the Bluetooth LE lamp with WoT abstraction., captionpos=b, label=lst:mit]
thing = consume(TD);
connect(thing);
status = thing.readProperty('power');
disconnect(thing);
\end{lstlisting}

\section{Related Work}
\label{section:RelatedWork}
The communication pattern of the Web of Things and the classification of device capabilities into properties, actions, and events have so far only been used for communication based on the network layer protocol IP. A number of different WoT protocol bindings for IP-based protocols have already been presented in the literature. The general method for developing protocol bindings was described by Mangas and Alonso \cite{GARCIAMANGAS2019235}. The introduced approach was applied to HTTP, Websockets, MQTT, and CoAP. However, protocol bindings have also been developed for more specialized protocols. For example, Sciullo et al. \cite{sciullo2020bringing} developed protocol bindings for the OPC UA and NETCONF protocols used in time-sensitive networks and presented additional vocabulary for semantic annotation within a Thing Description. There is also a W3C Editor's Draft for the Modbus \cite{Modbusbindings} protocol, which can be used to manage hardware in industrial settings. 

All these protocol bindings have in common that they are based on UDP or TCP and use the IP protocol. In contrast, we want to apply the WoT pattern to non-IP-based Bluetooth Low Energy communication. We focus especially on the application layer protocol GATT, which uses ATT in the transport layer and L2CAP in the network layer. 

node-wot implements already a codec which deserializes and serializes binary data\footnote{\url{https://github.com/eclipse/thingweb.node-wot/blob/master/packages/core/src/codecs/octetstream-codec.ts}} called \texttt{octet-stream}. The \texttt{octet-stream} codec takes the number of bytes (\texttt{length}), the sign (\texttt{signed}), the order of the bytes (\texttt{byteorder}) and the character set (\texttt{charset}) into account when operating on bytes. All parameters are passed as a comma-separated string. However, this information is not sufficient to adequately describe binary data transmitted by sensors. We therefore create a new codec as an extension to the existing one and introduce more relevant keywords (section \ref{abschnittCodec} and table \ref{tab:bdoOntolgy}). In addition, we do not want to pass the encoding or decoding information as a single string, but rather define each parameter using a separate RDF property in the Thing Description, which improves machine readability.

\section{Bluetooth Background}
\label{section:BackgroundOfBLE}
From the beginning, the primary objective of Bluetooth has been to exchange data wirelessly between devices. For this purpose, the Bluetooth Special Interest Group (SIG) created a standardized specification \cite{BluetoothLESpec}. Today there are two variants of the technology, Bluetooth Classic, also known as Bluetooth Basic Rate/Enhanced Data Rate (BR/EDR), and Bluetooth Low Energy (LE). Bluetooth Classic is used for direct data exchange between two devices, e.g. for audio streaming. Bluetooth LE was introduced with version 4.0 of the specification and was developed especially for small battery-operated devices.
Since most IoT devices are small, standalone devices, we assume that the relevance of Bluetooth LE in the IoT sector is far greater than that of Bluetooth Classic. 

\begin{figure}[h]
  \centering
  \includegraphics[width=6cm]{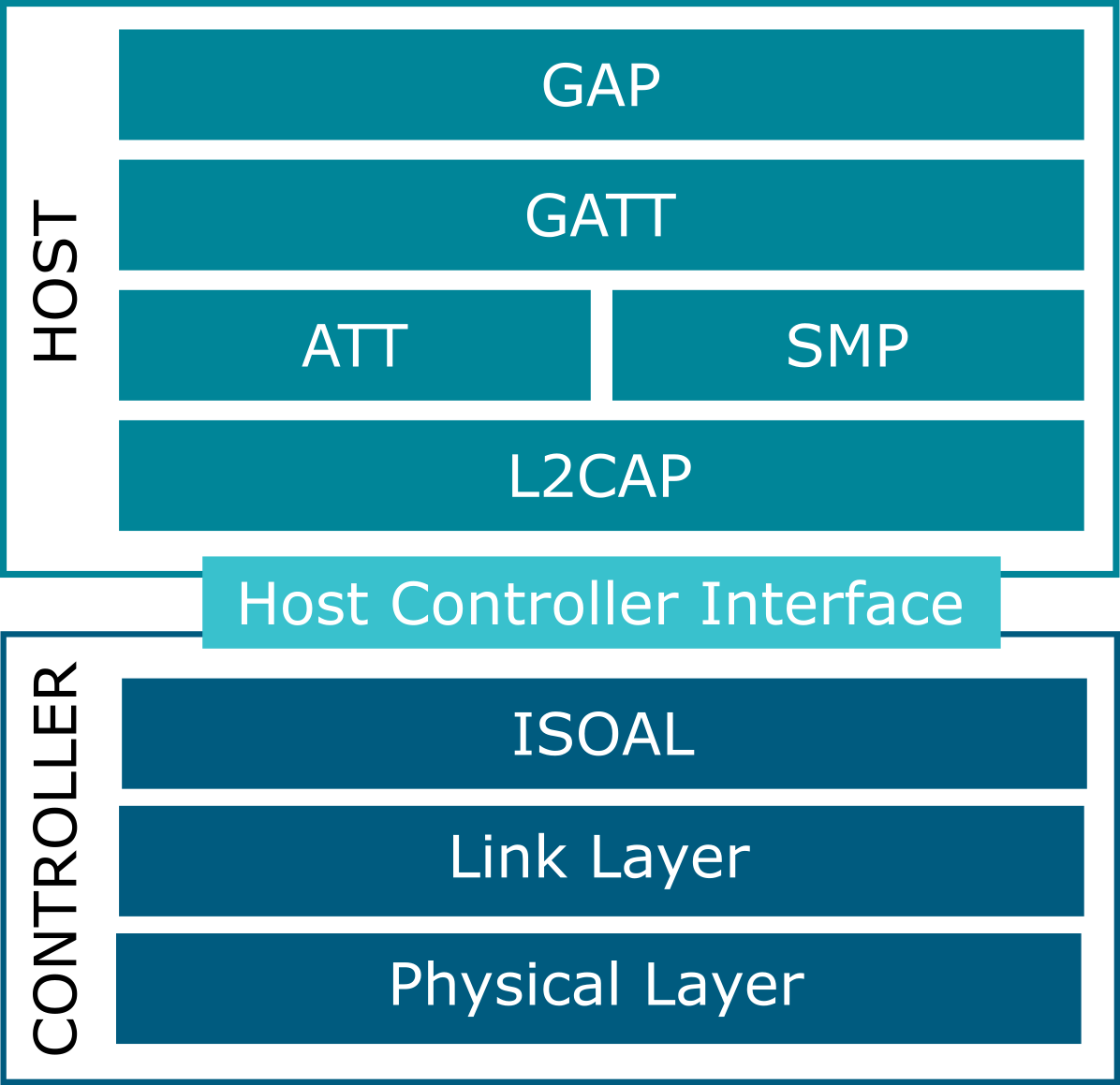}
  \caption{Each Bluetooth Low Energy device must implement the depicted protocol stack}
  \label{fig:BLEStack}
  \Description{The Bluetooth Low Energy protocol stack}
\end{figure}

The Bluetooth LE (BLE) protocol stack consists of two components, the host and the controller, which are connected via the host-controller interface (HCI), as shown in Figure \ref{fig:BLEStack}. The host is typically part of the operating system and the controller is embedded logic on the Bluetooth controller hardware. The highest communication level in BLE, i.e. the application layer, is formed by the so-called profiles, such as GAP and GATT. Profiles are based on the transport layer and use protocols such as ATT. The transport layer protocols in turn utilize L2CAP, a network layer protocol. The functionalities of the different profiles and protocols are presented in more detail below \cite{BluetoothLEPre}.

The \textbf{Generic Access Profile (GAP)} deals with high-level device discovery and connection establishment. For this purpose, the device roles Broadcaster and Observer, as well as Peripheral (i.e. server) and Central (i.e. client) are defined. The combination of Broadcaster and Observer is a connectionless communication type, whereas communication between Peripheral and Central is connection-oriented and thus forms a point-to-point connection. However, a Peripheral supports only a single connection at a time. A Central, on the other hand, is able to initiate and maintain multiple connections.

The \textbf{Attribute Protocol (ATT)} is used to read and write small data values (maximum 512 octets), but the protocol requires a connection for data exchange. Therefore, in the ATT environment, there are only the roles of a client and a server. The server organizes the data into attributes that are identified using so-called universally unique identifiers (UUIDs). The server can send responses to requests, receive data, and push asynchronous notifications to an ATT client. 

The \textbf{Generic Attribute Profile (GATT)} forms a superset of ATT, therefore a valid GATT client or server is also a valid ATT client or server. GATT specifies the structure of attributes contained on the server. GATT attributes are formatted in services, which in turn consist of characteristics. Characteristics hold a single data value and an indication of which GATT methods (see table \ref{tab:mappingWoT2BLE}) are allowed. The data value of a characteristic can optionally be described in more detail by descriptors to indicate the meaning of the value to a client. 
%All services, characteristics as well as descriptors are uniquely identified via 128 bit UUIDs. In general, a distinction is made between two types of UUIDs, the UUIDs standardized by the Bluetooth SIG and the freely selected custom UUIDs. The standardized UUIDs have a 16 bit short form with a 128 bit equivalent, these include for example a service for Battery and a characteristic for Battery Level. A list of all defined UUIDs is provided by the Bluetooth SIG. If there is no standardized UUID for a use case or an implementer does not want to use the standardized UUIds, custom UUIDs can be freely defined and assigned. 

The \textbf{Logical Link Control \& Adaptation Protocol (L2CAP)} has essentially three tasks, the protocol multiplexing where packets are forwarded to the different higher layers, the control of the data flow between various layers, and segmentation as well as reassembly of packets that exceed the Maximum Transmission Unit (MTU).

\section{Bluetooth LE for the Web of Things}
\label{section:BLEforWoT}
In the IoT area, devices must have low energy consumption and provide data in a structured way. This can be achieved with Bluetooth LE in combination with GATT. For this reason, we focus on the description of the GATT profile with the technologies and communication patterns introduced by the Web of Things.

\subsection{Design}
\label{chapter:design}
The W3C has defined requirements for creating new protocol bindings \cite{wotBinding}. The key considerations are the mapping of WoT operations to protocol methods, the definition of a URI scheme, and the specification of an appropriate media type.
These three points are covered in more detail below concerning our Bluetooth LE bindings.

\textbf{Default Mappings} In order to align and eventually integrate a new protocol into the WoT context, the required abstract WoT operations must first be mapped to the concrete operations of the new protocol. There are three categories of WoT operations. The first category is reading resources (e.g. \texttt{readproperty}), the second is writing resources (e.g. \texttt{writeproperty} or \texttt{invokeaction}), and the third is receiving notifications asynchronously (e.g. \texttt{subscribe}). The BLE GATT protocol has, among others, the methods \texttt{read}, \texttt{write} or \texttt{write-without-response}, and \texttt{notify}, which can be intuitively mapped to the three categories of WoT operations. However, a distinction must be made between the two GATT methods \texttt{write} and \texttt{write-without-response}. Both are capable of writing WoT resources, but the two methods differ in that \texttt{write} expects a confirmation message from the server after a write operation, while \texttt{write-without-response} requires no such confirmation. 
Thus the GATT method chosen depends on the implementation of the attribute in the GATT server. Therefore, in Table \ref{tab:mappingWoT2BLE} \texttt{write} and \texttt{write-without-response} are used interchangeably.

\begin{table}
  \caption{Mapping of well-known abstract WoT operations to concrete BLE GATT methods}
  \label{tab:mappingWoT2BLE}
  \begin{tabular}{cc}
    \toprule
    WoT Operation & BLE GATT Method\\
    \midrule
    \texttt{readproperty} & read \\
    \texttt{writeproperty}& write | write-w/o-response \\
    \texttt{invokeaction}&  write | write-w/o-response\\
    \texttt{readallproperties} & read \\
    \texttt{writeallproperties}& write | write-w/o-response\\
    \texttt{readmultipleproperties}& read\\
    \texttt{writemultipleproperties} & write | write-w/o-response\\
    \texttt{subscribeevent}& notify\\
    \texttt{unsubscribeevent}& notify\\
    \bottomrule
\end{tabular}
\end{table}

% Hier noch DataSchema hin?? 

\textbf{URI Scheme}
There is an expired Internet draft from the Internet Engineering Task Force (IETF) that addresses the issue of BLE URI schemes\footnote{\url{https://datatracker.ietf.org/doc/html/draft-bormann-t2trg-ble-uri-00}}. It introduces a URI scheme that describes both GAP operations such as scan and connect and GATT operations such as read and write characteristics. However, the focus is on gateway interaction rather than direct communication, and the approach has not been followed up. 
The draft proposes to get service links with a \texttt{GET} request on \textit{node/services} and with another \texttt{GET} request on \textit{service/characteristics} information about the characteristic can be obtained. 
Based on the considerations from the Internet Draft and the structuring of GATT attributes, we derived a custom URI scheme for GATT. Since we don't need information about services per se, but it is still advantageous for the implementation to know which characteristic is assigned to which service, we have defined the URI scheme in the form:
\begin{equation*}
  \textrm{\texttt{gatt://<deviceID>/<service>/<characteristic>}}
\end{equation*}
with the following meaning:
\begin{description}
   \item[\texttt{gatt}] Identification of the transfer protocol
   \item[\texttt{<deviceID>}] MAC address of the Bluetooth device
   \item[\texttt{<service>}] GATT service containing the characteristic
   \item[\texttt{<characteristic>}]  GATT characteristic to interact with
\end{description}
The introduced URI scheme is suitable for uniquely identifying resources on GATT servers, and allows users to interact with the desired GATT characteristic.

\textbf{Media Type}
\label{abschnittCodec}
In the \textit{forms} section of a Thing Description, a Media Type (also MIME Type or Content Type) can be specified that determines how sent data is encoded and received data is decoded. The encoding/decoding is done using a suitable codec. Currently, there is no fitting combination of content type and codec that meets all requirements for the binary data transmitted by Bluetooth LE. The closest is \textit{application/octet-stream}, but unfortunately, this codec only fulfills parts of our requirements and can not encode or decode all binary data transmitted via Bluetooth LE. Therefore, we have extended \textit{application/octet-stream} and its codec to meet our needs. This can be done by adding a new user-defined subtype for the 'application' type since all unrecognized subtypes are interpreted as \textit{application/octet-stream} and its codec by definition\footnote{\url{https://www.rfc-editor.org/rfc/rfc1521}}. 
For the Bluetooth LE bindings, we chose the new, non-standard subtype \textit{x.binary-data-stream} and defined an associated codec that interprets the binary data using a newly created vocabulary. 

\subsection{Vocabulary / Ontology}
\label{chapter:ontology}
The Thing Description is used to describe metadata and the interfaces of an existing Thing. However, to describe the possible interactions and transmitted binary data of a Bluetooth device, additional vocabulary is necessary. Therefore, we have developed two ontologies. One describes binary data in general, the other describes Bluetooth LE communication.

% Binary Data Ontology
\textbf{Binary Data Ontology} The key vocabulary terms of the Binary Data Ontology\footnote{\url{https://freumi.inrupt.net/BinaryDataOntology.ttl}} with the preferred prefix \texttt{bdo} are shown in Table \ref{tab:bdoOntolgy}. The bdo ontology is intended to provide maximum flexibility and describe all kinds of binary data. We want to use the bdo ontology to describe the data transmitted by Bluetooth LE devices, even those that do not comply with the Bluetooth standard (i.e. values of GATT characteristics should be in little-endian format as defined in Vol 3, Part G of \cite{BluetoothLESpec}). The terms of the vocabulary are in our use case only useful within the properties, actions, or events parts of a Thing Description because they contain information about the data that is needed in the codec associated with \textit{application/x.binary-data-stream}.
However, some Bluetooth devices expect not only the pure values but that a certain format is followed, for example, a certain start pattern followed by the actual value. For this case, the vocabulary provides the terms \texttt{bdo:pattern} and \texttt{bdo:variable}. With \texttt{bdo:pattern} a byte pattern encoded as a hex string with variables can be specified. The data type and further meta-information about the variables are then defined with \texttt{bdo:variable}. The user only has to provide the values of the variables, which are then inserted into the pattern before sending. The whole process is similar to URI variables. 
The terms \texttt{bdo:pattern} and \texttt{bdo:variable} can also be used for received data, in which case only the bytes at the specified positions are decoded by the codec.

\begin{table*}
  \caption{The key terms of the Binary Data Ontology}
  \label{tab:bdoOntolgy}
  \begin{tabular}{ccccc}
    \toprule
    Vocabulary term & Description & Assignment & Type & Default Value\\
    \midrule
    \texttt{bdo:bytelength} & Number of octets in the data & required & integer & None\\
    \texttt{bdo:signed} & Indicates if the data is signed & optional & boolean & false \\
    \texttt{bdo:endianess} & Byte order of the binary data & optional & string & bdo:littleEndian\\
     \texttt{bdo:offset} & Offset in number of octets & optional & integer & 0\\
    \texttt{bdo:scale} & Scale of received integer value & optional & float & 1.0\\
    \texttt{bdo:pattern} & Byte pattern of the binary data & optional & string & None\\
    \texttt{bdo:variable} & Description of the variables in  \texttt{bdo:pattern} & required, if \texttt{pattern} is used & --- & None \\
    \bottomrule
  \end{tabular}
\end{table*}

\begin{figure}[h]
  \centering
  \includegraphics[width=8cm]{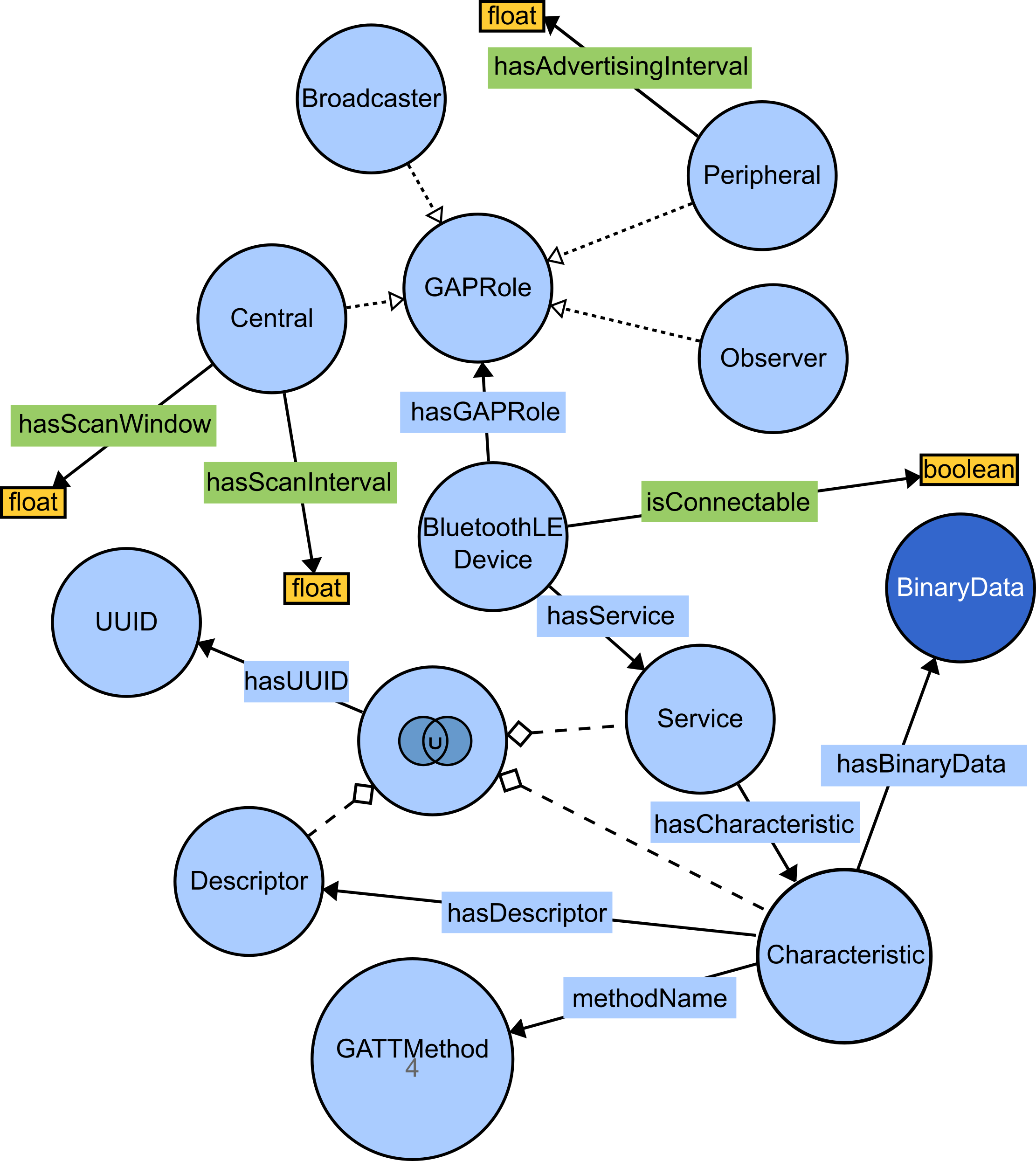}
  \caption{Classes and properties of the Simple Bluetooth Ontology}
  \label{fig:SBO}
  \Description{The Bluetooth Low Energy protocol stack}
\end{figure}

\textbf{Simple Bluetooth Ontology}
The communication and metadata of a Bluetooth Low Energy device is described using the Simple Bluetooth Ontology\footnote{\url{https://freumi.inrupt.net/SimpleBluetoothOntology.ttl}} with preferred prefix \texttt{sbo}. The sbo ontology provides classes and properties to describe the GAP role of a device, the structure of the data, the possible connection options, and low-level parameters of the Bluetooth LE link layer such as the length of the \texttt{scanWindow}, the \texttt{scanInterval} or the \texttt{advertisingInterval}. GATT Characteristics additionally have a property called \texttt{sbo:methodName} to specify allowed methods. In a Thing Description the \texttt{sbo:methodName} property is used in the \textit{forms} filed to indicate the required GATT method. Fig \ref{fig:SBO} shows an overview of the most relevant classes and properties of the ontology.

\subsection{Bluetooth LE Thing Description Example}
With the design decisions and the ontologies presented in section \ref{chapter:design} and section \ref{chapter:ontology}, it is possible to appropriately describe a Bluetooth LE device and its transmitted data. An example Thing Description of the Bluetooth LE lamp introduced in section \ref{section:Example} with a write property named \texttt{power} that can switch the light off or on is shown in listing \ref{lst:td}. 

Lines 9 to 14 contain metadata. From the metadata, we can tell that the Bluetooth LE lamp is a peripheral, allows connections, and broadcasts an advertisement every 50 milliseconds.

The 'power' characteristic of the Bluetooth LE lamp requires a start and end pattern in addition to the actual value to switch the light on or off. This is represented in the Thing Description of the \texttt{power} property via the term \texttt{bdo:pattern} (line 20). The pattern is provided in the form of a hex string and contains a variable named \textit{on} in curly brackets. Additional information about \textit{on} is provided by \texttt{bdo:variable} (line 21). From the annotation, it can be seen that \textit{on} needs to be an integer with allowed values of 0 or 1 encoded in a single byte. For the not explicitly defined terms of the bdo ontology, the default values are used. This means that the value of \textit{on} is converted to bytes in unsigned, little endian format without offset. 

In the \textit{forms} field, the method is specified as \texttt{sbo:write} (line 31), which corresponds to a normal write operation that expects a confirmation response. Also, \textit{application/x.binary-data-stream} is selected as the codec for the operation (line 32). 

To switch the Thing off or on, only a \texttt{writeproperty} operation to the \texttt{power} property containing the desired value of the variable \textit{on} must be executed. The actual code to switch the lamp on would look like this: $$\texttt{thing.writeproperty('power', \{on: 1\})}$$ The codec takes over the transformation of the integer value of the variable \textit{on} into bytes and inserts the byte value into the pattern before the data is sent to the Thing.

\begin{lstlisting}[float, language=json,firstnumber=1, caption=Example TD of a Bluetooth RGB controller using bdo:pattern and bdo:variable to switch on and off, captionpos=b, label=lst:td]
"@context": [
  'https://www.w3.org/2022/wot/td/v1',
 {
   'sbo': 'https://freumi.inrupt.net/SimpleBluetoothOntology.ttl#',
   'bdo': 'https://freumi.inrupt.net/BinaryDataOntology.ttl#',
   'rdf': 'http://www.w3.org/1999/02/22-rdf-syntax-ns#',
   'qudt': 'http://qudt.org/schema/qudt/' },],
 ...
'sbo:hasGAPRole': 'sbo:peripheral',
'sbo:isConnectable': true,
'sbo:hasGATTLayer': true,
'sbo:hasAdvertisingInterval': {
  'rdf:value': 50,
  'qudt:unit': 'qudt:MilliSEC'},
  
'properties': {
 'power': {
   'type': 'string',
   'format': 'hex',
   'bdo:pattern': '7e0004{on}00000000ef',
   'bdo:variable': {
    'on': {
     'type': 'integer',
     'minimum': 0,
     'maximum': 1,
     'bdo:bytelength': 1,
      } },
   'forms': [{
    'href': 'gatt://BE-58-30-00-CC-11/0000fff0-0000-1000-8000-00805f9b/0000fff3-0000-1000-8000-00805f9b34fb',
     'op': 'writeproperty',
     'sbo:methodName': 'sbo:write',
     'contentType': 'application/x.binary-data-stream',
    }], } },

\end{lstlisting}

\section{Using Bluetooth LE on Linux OS}
\label{section:BLEonLinux}
A large part of the high-end and low-end IoT operating systems is Linux-based \cite{bansal2020iot}. Therefore, we also want to focus on Linux. The official implementation of the Bluetooth protocol stack on Linux operating systems is BlueZ\footnote{\url{http://www.bluez.org/}}. The BlueZ stack is divided into two parts, one part is integrated into the official Linux kernel since version 2.4.6, and the other part is included in the user space and is available as the BlueZ package\footnote{\url{https://packages.debian.org/bullseye/bluez}}. 
The kernel handles low-level communication and security, the user space contains the central Bluetooth daemon \texttt{bluetoothd} and other tools provided by BlueZ like \texttt{btmgmt}\footnote{\url{https://manpages.debian.org/testing/bluez/btmgmt.1.en.html}}. The communication between the user level and the kernel is realized by sockets \cite{janc2016bluetooth}. 

There are three possible methods for how a user space application can interact and manage the Bluetooth controller. The first variant is the direct communication with the raw HCI interface, the second is the use of the Bluetooth management interface (mgmt API) and the third is the use of the BlueZ D-Bus interface. 

Commands and events sent and received directly through the HCI interface are the lowest level of communication with a Bluetooth adapter. The HCI is in the true sense a hardware abstraction and not well suited for creating high-level applications, because even for simple Bluetooth operations a large amount of code has to be written \cite{holtmann2006playing}. A detailed overview of the available HCI commands, functionalities, and events can be found in the Bluetooth specification in Vol. 4, Part E \cite{BluetoothLESpec}. Due to the difficulty of this API, we do not consider it any further.

The Bluetooth management interface was introduced in kernel 3.4 to address problems caused by the old direct communication method with HCI sockets, such as the execution of blocking operations or synchronization between the kernel and user space when sending commands \cite{mgmtAPI}. 
Available API calls to the Bluetooth management interface are, for example, the \texttt{Set Local Name Command} to change the local name of a controller, or the \texttt{Start Discovery Command} which starts the scan process and issues events when a device is found. These commands correspond internally to several HCI commands and are documented in the API specification\footnote{\url{https://git.kernel.org/pub/scm/bluetooth/bluez.git/tree/doc/mgmt-api.txt}}. The aforementioned command line tool \texttt{btmgmt} is based on the Bluetooth management interface.

In Linux systems, the so-called D-Bus\footnote{\url{https://www.freedesktop.org/wiki/Software/dbus/}} is used for interprocess communication. The Bluetooth daemon \texttt{bluetoothd} provides a BlueZ API for communication using the D-Bus system. The communication approach using the bus offers the advantage that almost all common programming languages can interact with the D-Bus and thus also with the Bluetooth stack. In addition, the interface is well documented\footnote{\url{https://git.kernel.org/pub/scm/bluetooth/bluez.git/tree/doc}} and is mentioned in the \textit{Bluetooth Technology for Linux Developers}\footnote{\url{https://www.bluetooth.com/bluetooth-resources/bluetooth-for-linux/}} course as the preferred variant. The command line tool \texttt{bluetoothctl}\footnote{\url{https://manpages.debian.org/stretch/bluez/bluetoothctl.1.en.html}} is based on the D-Bus API.

Since the D-Bus API has large documentation, is recommended by Bluetooth itself, and can be easily accessed in JavaScript via the D-Bus, we decided to use the D-Bus API for the implementation instead of the management API.

\begin{table*}[h]
  \caption{Time data of bluetoothctl (N=25)}
  \label{tab:bluetoothctlTime}
  \begin{tabular}{cccc}
    \toprule
    Device & Connect / ms & Disconnect / ms & read / ms \\
    \midrule
     BLE RGB Controller   & $837.76 \pm 23.14$ & $2379.96 \pm 48.05$& $100.91 \pm 4.29$ \\
      Arduino GATT Server & $1084.15\pm 37.21$  & $2326.93 \pm 58.77$   & $84.13 \pm 2.94$ \\
      Xiaomi Flower Care  & $2764.45 \pm 155.55$ & $2147.78 \pm 46.13$ & $81.28 \pm 2.79$ \\
    \bottomrule
\end{tabular}
\end{table*}

\begin{table*}[h]
  \caption{Time data of the BLE bindings (N=25)}
  \label{tab:bindingsTime}
  \begin{tabular}{cccc}
    \toprule
    Device & Connect / ms & Disconnect / ms & read / ms \\
    \midrule
     BLE RGB Controller   & $918.22 \pm 36.12$ & $2486.11 \pm 76.33$& $103.91 \pm 3.92$ \\
      Arduino GATT Server   & $1326.47 \pm 57.34$  & $2485.77 \pm 67.42$ & $86.89 \pm 2.48$  \\
     Xiaomi Flower Care  & $3206.37 \pm 255.54$ & $2263.17 \pm 10.23$ &  $84.51 \pm 3.52$ \\
    \bottomrule
\end{tabular}
\end{table*}

\section{Proof of Concept}
\label{section:Implementation}
% Vergleich auswahl Library? 
% Noble: gute performance, keine wartung, schlechte kompatibilität, events (notify) nur begrenzt möglich
% Node-ble: schlechtere performance, gewartet, gute kompatibilität
\subsection{Implementation}
To demonstrate the practicality of the theoretical approaches presented in the previous sections, we implemented Bluetooth Low Energy protocol bindings\footnote{\url{https://github.com/wintechis/Bluetooth-Bindings}}. We decided to use JavaScript as the programming language for the implementation of the Bluetooth LE bindings. The use of JavaScript ensures compatibility with node-wot, the reference implementation of the WoT Scripting API. 
Currently, only the client side is provided in the Bluetooth LE bindings, which allows communication with Bluetooth-enabled IoT devices of all kinds, but does not support creating custom GATT servers or exposing Bluetooth LE Things. 

We do not implement a new Bluetooth LE library from scratch to interact with the D-Bus API, but instead, use the node-ble\footnote{\url{https://github.com/chrvadala/node-ble}} package as the basis for Bluetooth LE communication in our bindings. node-ble is built on BlueZ and already uses the BlueZ D-Bus API to control connection establishment, termination, and all other necessary interaction options (\texttt{read}, \texttt{write}, \texttt{write-without-response}, and \texttt{notify}). In addition to the low-level communication methods of node-ble, we have implemented other higher-level operations for general connection management. The provided operations can be used to start and stop the scanning process or to select which devices should stay connected, which should be reconnected each time, and which should be disconnected.

\subsection{Limitations}
The implementation of the Bluetooth LE bindings also has limitations. Because the Bluetooth LE bindings are based on an already existing Bluetooth LE library for JavaScript called node-ble. Since this library requires BlueZ and the associated D-BUS API, the protocol bindings are only usable on Linux-based operating systems and require a small setup for communication with the DBus daemon. The close relationship between BlueZ and node-ble also means that the Bluetooth LE bindings have all the advantages and disadvantages of the BlueZ implementation.

Besides the requirement that BlueZ must be available, the Bluetooth LE bindings can only interact with GATT-based Bluetooth systems. Devices that broadcast their data via GAP advertisements cannot be used with the current implementation of the Bluetooth LE protocol bindings.

\section{Evaluation of the BLE bindings}
\label{section:Evaluation}
In this section, we evaluate the performance of the presented Bluetooth Low Energy bindings. To quantify the performance we first determined the latencies for different GATT operations using the WoT abstraction layer with our bindings and then performed the same GATT operations again with the well-established D-Bus-based tool \texttt{bluetoothctl} and compared the determined timing results with each other. 

All timing measurements were performed on an Intel NUC\footnote{\url{https://ark.intel.com/content/www/de/de/ark/products/126147/intel-nuc-kit-nuc8i5bek.html}} model NUC8i5BEK with Ubuntu 22.04 LTS (5.15.0-47) and BlueZ 5.65. Timings of the Bluetooth LE bindings were determined in Node.js with the command \texttt{performance.now()}\footnote{\url{https://nodejs.org/api/perf_hooks.html\#performancenow}}. 
We modified the source code of \texttt{bluetoothctl} in relevant places (\textit{/client/main.c} and \textit{src/shared/shell.c}) using the Linux system call \texttt{gettimeofday} to determine the execution time of various GATT operations. The modified version of the \texttt{bluetoothctl} software is available on GitHub\footnote{\url{https://github.com/FreuMi/bluez}}. Since both approaches are input and output-oriented, we measured elapsed real time rather than process time. 

The latency measurements were performed with three different IoT devices at a constant distance of one meter from the Central in an environment without any other Bluetooth LE devices. The three test devices are an RGB Light Controller\footnote{\url{https://github.com/arduino12/ble_rgb_led_strip_controller}} with an advertising interval of 50ms, an Arduino\footnote{\url{https://create.arduino.cc/projecthub/monica/getting-started-with-bluetooth-low-energy-ble-ab4c94}} with a custom GATT server and an advertising interval of 200 ms, and a Xiaomi Flower Care\footnote{\url{https://github.com/vrachieru/xiaomi-flower-care-api}} Sensor with an advertising interval of 2000 ms.

The timings were recorded for three operations, namely \texttt{connect}, \texttt{disconnect}, and \texttt{read}.
The time for \texttt{connect} covers the period between issuing the \textit{connect} command until it is theoretically possible to interact with characteristics. This includes finding the device, establishing a connection, and exploring the available services and characteristics.
The time interval for the \texttt{disconnect} operation starts after issuing the \textit{disconnect} command and ends as soon as the session has been successfully disconnected. 
The \texttt{read} operation starts again with the \textit{read} command and ends once the read value is available to the user.

A total of 25 measurements were performed for each operation and the arithmetic means and associated standard errors of the means were calculated (Table \ref{tab:bluetoothctlTime} and Table \ref{tab:bindingsTime}).

The data shows that the introduced Bluetooth LE bindings are on average about 16 percent slower than bluetoothctl when establishing a connection and about 6 percent slower when disconnecting. The higher latency is partly due to the programming language, bluetoothctl is written in C, and partly due to the fact that the Bluetooth LE bindings have the additional abstraction layer of the Web of Things.
The read operations, on the other hand, are almost identical and only differ by about 3 percent.

\section{Conclusion and Future Work}
\label{section:conclusion}
The Web of Things architecture's descriptive approach to create semantic interoperability between different incompatible networked devices is very promising. 

We have shown in this work that it is possible to apply the property-, action-, and event-based communication model of the Web of Things to non-IP-based protocols using  Bluetooth Low Energy interactions, in particular those involving servers whose attributes are formatted according to the GATT specification, as an example. We have designed ontologies that provide suitable vocabularies to describe metadata of Bluetooth Low Energy devices, various GATT methods, and the transmitted binary data, compared interaction possibilities with the Linux Bluetooth stack BlueZ, implemented protocol bindings with a matching codec for node-wot as a proof of concept, and evaluated the timing performance of the bindings. 
%This allowed us to demonstrate, that the WoT model in combination with the Thing Description can be suitable for non-Internet protocols in networks of embedded devices.

After laying the foundations for Bluetooth LE communication in the WoT context with this work, we want to extend the protocol bindings and ontologies in the next step. We plan to include not only GATT operations in our bindings but also other Bluetooth LE communication concepts and methods, such as listening to data that is sent via GAP advertisements and offering different Bluetooth LE security and encryption formats, which were out of scope in the Bluetooth LE bindings presented here. In addition, we want to use the technological basis and develop a WoT Servient-based gateway that can automatically consume Bluetooth LE devices described with a TD and expose them using an IP-based protocol.

\begin{acks}
This work was funded by the Bayerisches Verbundforschungsprogramm (BayVFP)
des Freistaates Bayern through the KIWI project (grant no. DIK0318/03)
\end{acks}

\bibliographystyle{ACM-Reference-Format}
\bibliography{src}

\end{document}